\documentclass[runningheads]{llncs}
\usepackage{filecontents}

\usepackage[T1]{fontenc}
\usepackage{graphicx}
\usepackage{hyperref}
\usepackage[baseline]{euflag}

\usepackage{listings}
\usepackage{url}
\begin{document}
\title{Correct orchestration of Federated Learning generic algorithms:
formalisation and verification in CSP\thanks{{\small\euflag} Funded by the European Union (TaRDIS, 101093006). Views and opinions expressed are however those of the author(s) only and do not necessarily reflect those of the European Union. Neither the European Union nor the granting authority can be held responsible for them.}}
\titlerunning{Correct orchestration of Federated Learning generic algorithms}
%
\author{Ivan Proki\'c\inst{1}\orcidID{0000-0001-5420-1527} \and
Silvia Ghilezan\inst{1,3}\orcidID{0000-0003-2253-8285} \and
Simona Ka\v sterovi\'c\inst{1}\orcidID{0000-0002-7161-3926}
	\and
Miroslav Popovic\inst{1}\orcidID{0000-0001-8385-149X}
	\and
Marko Popovic\inst{2}\orcidID{0000-0002-1957-0092}
	\and
	Ivan Ka\v stelan\inst{1}\orcidID{0000-0003-3417-7237}
 }
\authorrunning{I. Proki\'c et al.}
%
\institute{Faculty of Technical Sciences, University of Novi Sad, Novi Sad, Serbia\\
\email{\{prokic, gsilvia, simona.k, ivan.kastelan\}@uns.ac.rs\\
miroslav.popovic@rt-rk.uns.ac.rs} \\\url{http://www.ftn.uns.ac.rs/} \and
RT-RK Institute for Computer Based
Systems\\
\email{Marko.Popovic@rt-rk.com
} \and
Mathematical Institute of the Serbian Academy of Sciences and Arts, Belgrade, Serbia\\
\url{http://www.mi.sanu.ac.rs/}}
\maketitle              
\begin{abstract}
Federated learning (FL) is a machine learning setting where clients keep the training data decentralised and collaboratively train a model either under the coordination of a central server (centralised FL) or in a peer-to-peer network (decentralised FL). Correct orchestration is one of the main challenges. In this paper, we formally verify the correctness of two generic FL algorithms, a centralised and a decentralised one, using the CSP process calculus and the PAT model checker. The CSP models consist of CSP processes corresponding to generic FL algorithm instances. PAT automatically proves the correctness of the two generic FL algorithms by proving their deadlock freeness (safety property) and successful termination (liveness property). The CSP models are constructed bottom-up by hand as a faithful representation of the real Python code and is automatically checked top-down by PAT.


\end{abstract}
%
%
\section{Introduction}\label{sec: introduction}

Originally, {\em federated learning} (FL) was introduced by McMahan et al. \cite{McMahanMRHA17} as a decentralised approach to model learning that leaves the training data distributed on the mobile devices and learns a shared model by aggregating locally computed updates. Besides preserving local data privacy, FL is robust to the unbalanced and non-independent and identically distributed (non-IID) data distributions, and it reduces required communication rounds by 10–100x as compared to the synchronized stochastic gradient descent algorithm. Inspired by \cite{McMahanMRHA17}, Bonawitz et al. \cite{BonawitzIKMMPRS17} introduced an efficient secure aggregation protocol for federated learning, and Konecny et al. \cite{Konecny2017} presented algorithms for further decreasing communication costs. More recently, Bonawitz et al. \cite{BonawitzKMR22} and Perino et al. \cite{PerinoKLMK22} focused on data privacy.

Nowadays, there are many FL frameworks. The most prominent TensorFlow Federated (TFF) \cite{TensorFlow}, \cite{McMahan2023} and BlueFog \cite{BlueFog}, \cite{ying2021exponential} are well supported and accepted and they work well in cloud-edge continuum. However, they are not deployable to edge only, they are not supported on OS Windows, and they have numerous dependencies that make their installation far from trivial.

Recently, in 2021, Kholod et al. \cite{Kholod2021} made a comparative review and analysis of open-source FL frameworks for IoT, covering TensorFlow Federated (TFF) from Google Inc \cite{TensorFlow}, Federated AI Technology Enabler (FATE) from Webank’s AI department \cite{IGFLF}, Paddle Federated Learning (PFL) from Baidu \cite{DLP}, PySyft from the open community OpenMined \cite{openmined}, and Federated Learning and Differential Privacy (FL\&DP) framework from Sherpa.AI \cite{PrivPresAI}. They found out that application of these frameworks in the IoTs environment is almost impossible. So, developing a FL framework targeting smart IoTs in edge systems is still an open challenge.

More recently, in 2023, Popovic et al. proposed their solution to that challenge called Python Testbed for Federated Learning Algorithms (PTB-FLA) \cite{PopovicPKDG2023}. PTB-FLA was developed with the primary intention to be used as a FL framework for developing federated learning algorithms (FLAs), or more precisely as a runtime environment for FLAs. The word “testbed” in the name PTB-FLA that might be misleading was selected by ML \& AI developers in TaRDIS project \cite{TaRDIS} because they see PTB-FLA as an “algorithmic” testbed where they can plugin and test their FLAs. Note that PTB-FLA is neither a system testbed, such as the one that was used for testing the system based on PySyft in \cite{ChengXue2021}, nor a complete system such as CoLearn \cite{Feraudo2020} and FedIoT \cite{Zhang2021} (for more elaborated comparison with CoLearn and FedIoT see Section I.A in \cite{PopovicPKDG2023}).

PTB-FLA is written in pure Python to keep the application footprint small so to fit to IoTs, and to keep installation as simple as possible (with no external dependencies). PTB-FLA supports both centralised and decentralised FLAs. The former is as defined in \cite{McMahanMRHA17}, whereas the latter are generalized such that each process (or node) alternatively takes server and client roles from \cite{McMahanMRHA17} or more precisely, it switches roles from server to client and back to server.

PTB-FLA enforces a restricted programming model, where a developer writes a single application program, which is later instantiated and launched by the PTB-FLA launcher as a set of independent processes, and within their application program, a developer only writes callback functions for the client and the server roles, which are then called by the generic federated learning algorithms hidden inside PTB-FLA.

So far, PTB-FLA usage has been illustrated and validated by three simple examples in \cite{PopovicPKDG2023}, but PTB-FLA has not been formally verified. In this paper, we formally verify the correctness of two generic FL algorithms, a centralised and a decentralised one, using the CSP process calculus and the PAT model checker, in a process with two phases.

In the first phase, we construct by hand CSP models of the generic centralised and decentralised FLAs as faithful representations of the real Python code. We construct these models in a bottom-up fashion in two steps. In the first step, we construct processes corresponding to generic FL algorithm instances, and in the second step, we construct the system model as an asynchronous interleaving of $n$ FL algorithm instances.

In the second phase, we formally verify CSP models constructed in the previous phase in two steps. In the first step, we formulate desired system properties, namely deadlock freeness (safety property) and successful FLA termination (liveness property). We formulate the latter property in two equivalent forms (reachability statement and always-eventually LTL formula). In the second step, we use PAT to automatically prove formulated verification statements.

The main contributions of this paper are: (i) the CSP models of the generic centralised and decentralised FLAs, (ii) the formulations of generic centralised and decentralised FLAs properties. To the best of our knowledge, this is the first paper that formally verifies decentralised FLAs.

The rest of the paper is organized as follows. Section~\ref{sec: related work} presents closely related work. Section \ref{sec: ptb_fla} presents the PTB-FLA overview, Section \ref{sec: formalization} presents PTB-FLA formalization, Section \ref{sec: verification} presents PTB-FLA verification, and Section \ref{sec: conclusion} concludes the paper.

\subsection{Short Discussion of Closely Related Work}\label{sec: related work}

While tools for decentralised ML (DML), especially FL, are starting to flourish, many are not flexible and portable enough to experiment with novel processors, not fully connected network topologies, and asynchronous schemes. To overcome these limitations, Mittone et al. use the formal language RISC-pb2l to describe distributed FL workloads and to map them to the FastFlow parallel programming library \cite{mittone2023}. We consider this approach as orthogonal to our work because it targets parallel and distributed processing composition and optimization whereas our work targets formal verification of system correctness, i.e. proving desired system properties.

Multiparty Asynchronous Session Types (MPST) is a class of behavioural types tailored for describing distributed protocols relying on asynchronous communications. Hu and Yoshida extended MPST in \cite{HuY17} with explicit connection actions to support protocols with optional and dynamic participants. Although these extended MPST enabled modelling and verification of some protocols in cloud-edge continuum \cite{SimicPDSM21}, we could not use them to model the generic centralised and decentralised FLAs, because we could not express arbitrary order of message arrivals that take place at an FLA instance.

The design of robust protocols for coordination of peer-to-peer systems is difficult because it is hard to specify and reason about their global behaviour. Recently, Kuhn et al. presented an approach in \cite{kuhn2023} where a so-called swarm protocol is a global system specification, whereas swarm protocol projections to machines are local specifications of peers. 
They claim that swarms are deadlock free, but liveness is not guaranteed in their theory. We find this approach interesting and in our future work we plan to investigate whether it would be feasible for our generic FLAs. 
At present, we 
identify some of the differentiating points between \cite{kuhn2023} and our work: $(i)$ in their approach communication of peers is conducted through a shared log instead of point-to-point message passing; 
$(ii)$ they model 
peers using finite state automata, while we use (CSP) processes;
$(iii)$ they model protocols in the style of MPST via top-down approach (projecting global type onto peers to obtain local type specification) while we only write local processes specifications, that we ensemble together to obtain global protocol behaviour; 
$(iv)$ they use TypeScript language and develop tools to check protocol conformance at runtime through equivalence testing, whereas our protocols are written in Python language, 
modeled in CSP, and we use PAT  to prove deadlock freeness and liveness. 

\section{Generic Federated Learning Algorithms: PTB-FLA Overview}\label{sec: ptb_fla}

This section presents the PTB-FLA overview. The term {\em PTB-FLA system} refers to a system based on PTB-FLA. The next three subsections present the PTB-FLA system architecture, the PTB-FLA API, and the PTB-FLA system operation, respectively.

\begin{figure}[t]
\centering
\includegraphics[scale=0.65]{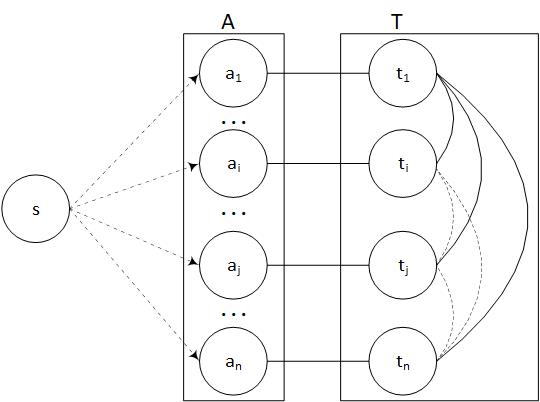}
\caption{Block diagram of the PTB-FLA system architecture.}\label{fig: system architecture}
\end{figure}

\subsection{PTB-FLA System Architecture}

The PTB-FLA system architecture is composed of the application launcher process s, the distributed application $A = \{a_1, a_2, \ldots , a_n \}$, and the distributed testbed $T = \{t_1, t_2, \ldots , t_n\}$, see Fig.~\ref{fig: system architecture}, where $a_i$ is an application program instance, $t_i$ is a testbed instance, and $n$ is the number of instances in both $A$ and $T$. The distributed application $A$ uses the distributed testbed $T$ to execute the distributed algorithm, which is specified by the callback functions within the application program. PTB-FLA supports both centralised and decentralised federated learning algorithms by providing the API functions that implement the generic centralised algorithm and the generic decentralised algorithm, named fl\_centralised and fl\_decentralised, respectively.

A particular distributed federated learning algorithm is executed as follows. Each instance $a_i$ prepares its input data based on the command line arguments (including the identification $i$, the number of instances $n$, etc.) and then calls the desired generic API function on its testbed instance $t_i$.

The testbed instance $t_i$ in turn plays its role in the generic algorithm by exchanging messages with other testbed instances and by calling the associated callback function at the right point of the generic algorithm. The communication graph of testbed instances either takes the form of a star in case of a centralised algorithm (see solid edges connecting the server $t_1$ and the clients $t_2$ to $t_n$ in Fig.  \ref{fig: system architecture}), or the form of a clique in the case of a decentralised algorithm (see solid and dashed edges connecting all the testbed instances in Fig. \ref{fig: system architecture}).

\begin{figure}[t]
\centering
\includegraphics[scale=0.65]{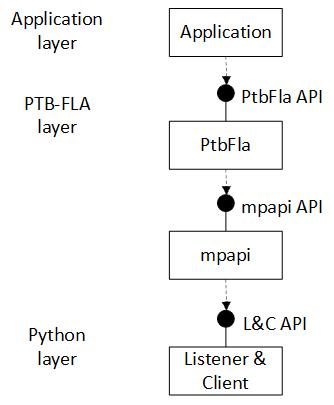}
\caption{UML class diagram of the PTB-FLA system architecture.}\label{fig: UML class diagram}
\end{figure}

Fig. \ref{fig: UML class diagram} shows the simplified UML class diagram of a PTB-FLA system. The PTB-FLA system architecture comprises three layers: the distributed application layer, the PTB-FLA layer (comprising the class PtbFla in the module ptbfla and the module mpapi) in the middle, and the Python layer at the bottom. The application module uses the PtbFla to create or destroy a testbed instance and to conduct its role in the distributed algorithm execution by calling the API function fl\_centralised or the API function fl\_decentralised.

The API functions fl\_centralised and fl\_decentralised, within an instance $t_i$, use the module mpapi (mpapi is the abbreviation of the term {\em message passing API}) to communicate with other instances. The module mpapi in turn instantiates the Python multiprocessing classes Listener and Client to create the mpapi server and the mpapi client, which are hidden with the module mpapi and provide reliable TCP connections among testbed instances.

\subsection{PtbFla API}\label{sec:ptb-fla_api}

The PtbFla API offers the constructor, two generic FLAs, and the destructor:

\begin{itemize}
\item PtbFla($noNodes, nodeId, flSrvId=0$)
\item $ret$ fl\_centralised($sfun, cfun, ldata, pdata, noIters=1$)
\item $ret$ fl\_decentralised($sfun, cfun, ldata, pdata, noIters=1$)
\item PtbFla()
\end{itemize}

The arguments are as follows: $noNodes$ is the number of nodes (or processes), $nodeId$ is the node identification, $flSrvId$ is the server id (default is $0$; this argument is used by the function fl\_centralised), $sfun$ is the server callback function, $cfun$ is the client callback function, $ldata$ is the initial local data, $pdata$ is the private data, and $noIters$ is the number of iterations that is by default equal to $1$ (for the so called one-shot algorithms), i.e. if the calling function does not specify it, it will be internally set to $1$. The return value $ret$ is the node final local data. Data ($ldata$ and $pdata$) is application specific.

Typically, $ldata$ is a machine learning model, whereas $pdata$ is a training data that is used to train the model. Normally, the testbed instances only exchange $ldata$ and they never send out $pdata$ (that is how they guarantee the training data privacy). The $pdata$ is only passed to callback functions within the same process instance to immediately set them in their working context.

\begin{figure}[t]
\centering
\includegraphics[scale=0.65]{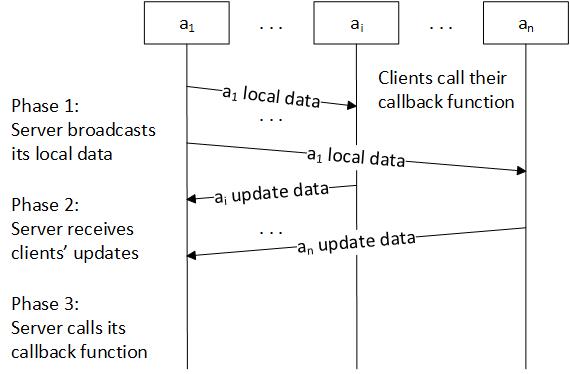}
\caption{The generic centralised one-shot FLA execution.}\label{fig: generic centralized}
\end{figure}

\subsection{PTB-FLA Operation}\label{sec:ptb-fla_algs}

This subsection provides an overview of the PTB-FLA operation by presenting the two most important scenarios: the generic centralised and decentralised one-shot FLA executions, respectively.

The generic centralised one-shot FLA has three phases, see Fig. \ref{fig: generic centralized} (here $a_1$ is the server and $a_i$, $i = 2, \ldots, n$, are the clients). In the first phase, the server broadcasts its local data to the clients, which in their turn call their callback function to get the update data and store the update data locally. In the second phase, the server receives the update data from all the clients (in any order, caused by arbitrary delays), and in the third phase, the server calls its callback function to get its update data (i.e. aggregated data) and stores it locally. Finally, all the instances return their new local data as their results.

Unlike the generic centralised FLA that uses the single field messages carrying data, the generic decentralised FLA uses the three field messages carrying: the messages sequence number (i.e. the phase number), the message source address (i.e. the source instance network address), and the data (local or update).

\begin{figure}[h!]
\centering
\includegraphics[scale=0.65]{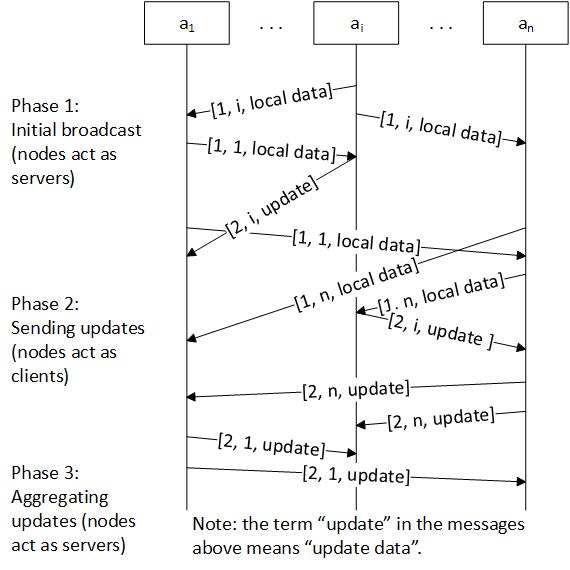}
\caption{The generic decentralised one-shot FLA execution.} \label{fig: generic decentralized }
\end{figure}

The generic decentralised one-shot FLA has three phases, see Fig. \ref{fig: generic decentralized }. In the first phase, each instance acts as a server, and it sends its local data to all its neighbours. These messages have the sequence number $1$, each instance sends $(n-1)$ such messages and is also the destination for $(n-1)$ such messages.

In the second phase, each instance acts as a client, and it may receive either a message with the sequence numbers $1$ or $2$. In the latter case, it just stores it in a buffer for later processing in the third phase, whereas in the former case, it calls the client callback and sends the update data in the reply to the message source. Note that during the second phase, the instance does not update its local data, it just passes the update data it got from the client callback function.

Since messages are sent asynchronously, they may be received in any order. Fig. \ref{fig: generic decentralized } shows a scenario where the instance $a_1$ receives the messages in the messages sequence $1-2-1-2$, which is out of the phase order, whereas the instances $a_i$ and $a_n$ receive the messages in the sequence $1-1-2-2$, which is in the phase order. However, by using the abovementioned buffering, the instance $a_1$ postpones processing of the phase $2$ messages until the third phase.

The second phase is completed after the instance received and processed all $2(n - 1)$ message. In the third phase, each instance again acts as a server, and it calls the server callback function to get its update data (e.g., aggregated data) and stores it locally. Finally, all the instances return their new local data as their results.

\section{CSP Formal Models}\label{sec: formalization}

In this section we use CSP process calculus to obtain a formal specification of the communication layer of our PTB-FLAs. 
The CSP provides modeling of the concurrency primitives as follows:
\begin{itemize}
\item the system components are CSP \emph{processes};
\item communication between the system components is performed through the \emph{communication channels};
\item the system of parallel processes communicating asynchronously (i.e. without barrier synchronization) is assembled via \emph{interleaving} of the CSP processes.
\end{itemize} 

The rest of the section is organized as follows: Section~\ref{sec:modeling_centralized_algorithm} presents the model for our centralised algorithm and Section~\ref{sec:modeling_decentralized_algorithm}  presents the model for the decentralised algorithm.

\subsection{Modeling centralised algorithm}\label{sec:modeling_centralized_algorithm}

\begin{figure}
\begin{lstlisting}[frame=single]
// PTB-FLA
enum {False, True};
#define NoNodes 3;
#define FlSrvId 2;
var ldataArr[NoNodes];
var pdataArr[NoNodes];
var terminated;
channel server2client[NoNodes-1] 1;  
channel clients2server NoNodes-1;    

FlCentralised(noNodes, nodeId, flSrvId, ldata, pdata) =
  if(nodeId == FlSrvId) {
    CeServer(noNodes, nodeId, flSrvId, ldata, pdata)
  } else {
    CeClient(noNodes, nodeId, flSrvId, ldata, pdata)
  };

CeServer(noNodes, nodeId, flSrvId, ldata, pdata) =
  {terminated = False} ->
  CeBroadcastMsg(0, noNodes, nodeId, ldata);
  CeRcvMsgs(0, noNodes-1);
  {terminated = True} -> Skip;

CeBroadcastMsg(id, noNodes, nodeId, ldata) =
  if(id != nodeId) {
    server2client[id]!ldata -> Skip
  };
  if(id < noNodes-1) {
    CeBroadcastMsg(id+1, noNodes, nodeId, ldata)
  };

CeRcvMsgs(i, noMsgs) =
  if(i < noMsgs) {
    clients2server?update -> CeRcvMsgs(i+1, noMsgs)
  };

CeClient(noNodes, nodeId, flSrvId, ldata, pdata) =
  server2client[nodeId]?srvLdata ->
  clients2server!ldata+srvLdata ->
  Skip;

SysCentralised() = 
  |||nodeId:{0..NoNodes-1}
  @FlCentralised(NoNodes, 
                 nodeId, 
                 FlSrvId, 
                 ldataArr[nodeId], 
                 pdataArr[nodeId]);
\end{lstlisting}
\caption{CSP model for centralised algorithm.}
\label{fig:CSP_centralized}
\end{figure}

Figure~\ref{fig:CSP_centralized} shows a CSP model for our centralised algorithm. 
Lines 2-3 define number of nodes  (\lstinline{NoNodes}) 
(indexed with $0, 1, 2, \ldots$) with the server (\lstinline{FlSrvId}) having the largest index, and other nodes being clients.  
We remark we could set here the index of the server node with the smallest index (as it is in Section~\ref{sec:ptb-fla_algs}), but this would in fact make our model less intuitive because of the channel manipulation (as explained bellow). 
Lines 4-5 define arrays of local data \lstinline{ldata} and private data \lstinline{pdata} - one per each node. 
The communication channels are defined in lines 8-9. 
The array of channels \lstinline{server2client}    - one per each client (hence, \lstinline{NoNodes}$-1$ channels) are used for the server broadcast of their local data to the clients (one channel per client). Notice that the indexes of array elements are generated starting with $0$, hence the channel index indicates the index of the client node.   
Since we consider one-shot algorithm the server sends their local data only once, hence the channels are specified to have FIFO buffers of size $1$. 
Channel \lstinline{clients2server} is used in the second phase of our algorithm, i.e. for clients replying to the server with the update data. The FIFO size of this channel is \lstinline{NoNodes}$-1$, since all clients reply with a single update. 

Lines 11-16 define a generic node as a CSP process with parameters of the number of nodes, identification of the node, index of the server, their local and private data. 
We remark that parameters $sfun, cfun$, and $noIters$, also present in fl\_centralised (cf. Section~\ref{sec:ptb-fla_api}), were considered out of the scope for this model. 
Based on the node index the process proceeds as the server node \lstinline{CeServer} or as one of the client nodes \lstinline{CeClient}. 

The server node is modeled in lines 18-22. The process first checks if it is terminated: if not it performs the broadcasting of the local data via \lstinline{CeBroadcastMsg} (i.e. it enters the phase $1$, cf. Figure~\ref{fig: generic centralized}), then proceeds to phase $2$ by receiving updates  via \lstinline{CeRcvMsgs}. 
The successful termination is modeled with \lstinline{Skip}. 
The broadcasting of server's local data \lstinline{CeBroadcastMsg} is defined in lines 24-30. 
The server sends ldata on channels \lstinline{server2clients[id]} (if \lstinline{id} is not their own index), and then recursively calls itself with index increased by 1 - if the index is less then \lstinline{noNodes}$-1$. 
Since \lstinline{CeServer} passes \lstinline{id} to \lstinline{CeBroadcastMsg} to be 0, the server will send the local data to all the clients exactly once. 
Once the broadcast is done, the server starts receiving clients' updates on channel \lstinline{clients2server} as defined with  \lstinline{CeRcvMsgs} in lines 32-35.

The client process is defined with \lstinline{CeClient} in lines 37-40. 
The client with index \lstinline{nodeId} first receives server's local data on channel \lstinline{server2client[nodeId]}, and then replies updated server's local data with its own local data (here for simplicity modeled with addition) on channel \lstinline{clients2server}, after which client process successfully terminates.

The system consisting of \lstinline{NoNodes}$-1$  clients and a single server is then modeled as the interleaving of the \lstinline{FlCentralised} processes (lines 42-48), since all processes but one indexed  \lstinline{FlSrvId} are instantiated as clients (and the one indexed \lstinline{FlSrvId} is instantiated as a server).

\subsection{Modeling decentralised algorithm}\label{sec:modeling_decentralized_algorithm}

\begin{figure}
\begin{lstlisting}[frame=single]
// PTB-FLA
enum {False, True};
#define NoNodes 3;
var ldataArr[NoNodes];
var pdataArr[NoNodes];
var terminated;
channel tonode[NoNodes] 2*(NoNodes-1);  
channel buffer[NoNodes] NoNodes-1; 

FlDecentralised(noNodes, nodeId, ldata, pdata) =
  {terminated = False} ->
  DeBroadcastMsg(0, noNodes, nodeId, ldata);
  DeRcvMsgs(0, noNodes, nodeId, ldata);
  DeRcvMsgs2(0, noNodes, nodeId);
  {terminated = True} -> Skip;

DeBroadcastMsg(id, noNodes, nodeId, ldata) =
  if(id != nodeId) {
    tonode[id]!1.nodeId.ldata -> Skip
  };
  if(id < noNodes-1) {
    DeBroadcastMsg(id+1, noNodes, nodeId, ldata)
  };

DeRcvMsgs(i, noNodes, nodeId, ldata) =
  if(i < 2*noNodes-2) {
    tonode[nodeId]?phase.from.nodeldata -> 
    if(phase == 1){
    tonode[from]!2.nodeId.ldata+nodeldata -> 
    DeRcvMsgs(i+1, noNodes, nodeId, ldata)
    } else {
    buffer[nodeId]!phase.from.nodeldata ->
    DeRcvMsgs(i+1, noNodes, nodeId, ldata)
    }
  };

DeRcvMsgs2(i, noNodes, nodeId) =
  if(i < noNodes-1) {
    buffer[nodeId]?phase.from.update -> 
    DeRcvMsgs2(i+1, noNodes-1, nodeId)
  };

SysDecentralised() = 
  |||nodeId:{0..NoNodes-1}
  @FlDecentralised(NoNodes, 
                   nodeId, 
                   ldataArr[nodeId], 
                   pdataArr[nodeId]);
\end{lstlisting}
\caption{CSP model for decentralised algorithm.}
\label{fig:CSP_decentralized}
\end{figure}

The CSP model for our decentralised algorithm is given in Figure~\ref{fig:CSP_decentralized}. 
Albeit more complex than the centralised one, the decentralised algorithm yields a slightly simpler CSP model. 
The reason is that all nodes in the system have the same behaviour. In phase $1$ all nodes behave as servers broadcasting their local data to all other nodes, which in turn update the data and return an answer in phase $2$ (corresponding to phases given in Figure~\ref{fig: generic decentralized } in Section~\ref{sec:ptb-fla_algs}). 
All the nodes receive messages from all other nodes as they arrive, but first process the messages from phase $1$ and only then deals with the messages from the phase $2$. 
We model this behaviour with assigning two channels to each process (i.e. node). 
One channel is for receiving messages from other processes, called \lstinline{tonode}, with buffer of size \lstinline{2*(NoNodes-1)} (line 7), since the node will receive messages from all other nodes from both phases. 
The other channel assigned to node, called \lstinline{buffer} (line 8), serves only for storing messages from the second phase while all messages from the first phase are processed - later in phase $3$ the same node will read those messages. 
Hence, the buffer size of these channels are \lstinline{NoNodes-1}. 

The node processes are defined with \lstinline{FlDecentralised} in lines 10-15. 
Process first broadcasts their local data with \lstinline{DeBroadcastMsg} (defined in lines 17-23) - which behaves in the same way as \lstinline{CeBroadcastMsg} in the centralised algorithm (cf. Figure~\ref{fig:CSP_centralized}), except that the sent messages now contain not only field for local data of the node, but also fields marking the phase (here $1$) and the node's index (that the receiving node uses for the reply in phase $2$). 
The node then proceeds with receiving messages from all other nodes with \lstinline{DeRcvMsgs}, and finally (phase $3$) process the messages from the second phase with \lstinline{DeRcvMsgs2}.

\lstinline{DeRcvMsgs} is given in lines 25-35. 
Here we deviate from the centralised algorithm: node receives all messages from both phases from the other nodes and then performs an analysis on the phase of the received message. 
If the phase is $1$, the node replies updated data to \lstinline{from} they received message in the first place, marking the phase of the message $2$. 
If, on the other hand, the phase is $2$, the node stores the message to their own channel \lstinline{buffer[nodeId]}. 
Once the node process all messages from phase $1$ (and buffers all messages from phase $2$), \lstinline{DeRcvMsgs2} (lines 37-41) is used to read from the \lstinline{buffer[nodeId]}, which behaves in the same way as \lstinline{CeRvcMsgs} from the centralised algorithm (cf. Figure~\ref{fig:CSP_centralized}). 

The system of \lstinline{NoNodes} nodes is finally modeled as the interleaving of the
 \lstinline{FlDecentralised} processes in lines 43-48.
 
\section{Formal Verification in PAT}\label{sec: verification}

The correctness of our CSP models is automatically checked by PAT, that supports the system analysis in two ways: simulation and model checker.  We have used the latter one. 

\begin{figure}[t]
\begin{lstlisting}[frame=single]
// ...
// CSP model for centralised algorithm
// ...

#assert SysCentralised() deadlockfree;
#define Terminated (terminated == True);
#assert SysCentralised() reaches Terminated;
#assert SysCentralised() |= []<> Terminated;
\end{lstlisting}
\caption{Verifying centralised algorithm.}
\label{fig:verfying_centralized}
\end{figure}

The correctness of our centralised and decetralised algorithms is verified by proving the deadlock freeness (safety property) and successful termination (liveness property). The properties about algorithms are stated in the form of queries, called {\em assertions}, which are checked by PAT.  The assertions that formally verify the correctness of our centralised algorithm are shown in Figure~\ref{fig:verfying_centralized}.

The assertion  
 given in line $5$ of  Figure~\ref{fig:verfying_centralized} claims that the centralised algorithm is deadlock free. PAT model checker performs Depth-First-Search or Breath-First-Search algorithm to check if the assertion is true. It explores unvisited states until a non-terminated state with no further move---
called a \emph{deadlock state}, is found or all states have been visited. 

The assertion 
given in line $7$ of Figure~\ref{fig:verfying_centralized} claims that the centralised algorithm reaches a terminated state. This assertion is checked by performing Depth-First-Search algorithm. PAT model checker repeatedly explores all unvisited states until it finds a state at which the condition \lstinline{Terminated} is satisfied or  it visits all the states. The condition \lstinline{Terminated} is a proposition defined as a global definition (line $6$ in Figure~\ref{fig:verfying_centralized}). 

PAT supports the full syntax of the linear temporal logic (LCL), which is used in the last assertion of Figure~\ref{fig:verfying_centralized} 
that claims our centralised algorithm satisfies formula \lstinline{[]<> Terminated}. The modal operator \lstinline{[]} reads as ''always'' and the operator \lstinline{<>} reads as ''eventually'', so 
statement asserts our centralised algorithm always eventually reaches the terminated state.

\begin{figure}[t]
\begin{lstlisting}[frame=single]
// ...
// CSP model for decentralised algorithm
// ...

#assert SysDecentralised() deadlockfree;
#define Terminated (terminated == True);
#assert SysDecentralised() reaches Terminated;
#assert SysDecentralised() |= []<> Terminated;
\end{lstlisting}
\caption{Verifying decentralised algorithm.}
\label{fig:verifying_decentralized}
\end{figure}

The proof of correctness of our decentralised algorithm is given in Figure~\ref{fig:verifying_decentralized}, and follows the same explanations given for the centralised one.

\section{Conclusion}\label{sec: conclusion}
In this paper, we formally verified the correctness of two generic FL algorithms, a centralised and a decentralised one, using the CSP process calculus and the PAT model checker. The CSP models are constructed bottom-up by hand as a faithful representation of the real Python code and their correctness (safety and liveness) are automatically checked top-down by PAT.

The main contributions of this paper are: 
\begin{itemize}
	\item the CSP models of the generic centralised and decentralised FLAs,
	\item  the formulations of generic centralised and decentralised FLAs properties. To the best of our knowledge, this is the first paper that formally verifies decentralised FLAs.
\end{itemize} 
The main limitations of this paper are: 
\begin{itemize}
	\item we implicitly assumed that callback functions are terminating (i.e., have termination property),
	\item  we did not model any ML\&AI processing within the callback functions and therefore were unable to address the properties of the corresponding information flows and output results, such as privacy of information flows, understandability/interpretability of the resulting models, etc.
\end{itemize} 

In our future work, we may try to address some of the latter limitations mentioned above.

\bibliographystyle{splncs04}
 \bibliography{references}
\end{document}